\documentclass[twocolumn,aps,prb,raggedbottom,nobalancelastpage,amssymb,superscriptaddress]{revtex4}

\usepackage{amsmath}
\usepackage{amssymb}
\usepackage{amsfonts}
\usepackage{dsfont}
\usepackage{graphicx}
\usepackage{bm}
\usepackage{color}
\usepackage{appendix}
\usepackage{epsfig}

\begin{document}
\title{Sharp correlations in the ARPES spectra of strongly disordered topological boundary modes}

\author{Zohar Ringel}
\affiliation{Theoretical Physics, Oxford University, 1, Keble Road, Oxford OX1 3NP, United Kingdom.}


\begin{abstract}
Data from angle resolved photo-emission spectroscopy (ARPES) often serves as a smoking-gun evidence for the existence of topological materials. It provides the energy dispersion curves of the topological boundary modes which characterize these phases. Unfortunately this method requires a sufficiently regular boundary such that these boundary modes remain sharp in momentum space. Here the seemingly random data obtained from performing ARPES on strongly disordered topological insulators and Weyl semimetals is analyzed theoretically and numerically. Expectedly the disorder averaged ARPES spectra appear featureless. Surprisingly however, correlations in these spectra between different energies and momenta reveal delta-sharp features in momentum space. Measuring such correlations using nano-ARPES may verify the topological nature of the suggested weak topological insulator ($Bi_{14} Rh_3 I_9$) which thus far was not studied using ARPES due to the rough nature of its metallic surfaces. 
 \end{abstract}


\maketitle

Topological insulators (TIs) and semimetals have attracted much attention in recent years due to their novel bulk and surface properties~\cite{RMP_TI,RMP_TI2,Arcs2015}. Their bulk band structure has certain twists characterized by robust topological numbers. These abstract topological numbers become very much real on the boundary of the material where exotic metallic phases emerge. For instance, the surface of a 3D Strong topological insulator can host a single Dirac cone with momentum and spin locked together. Similarly unconventional are the open Fermi arcs occurring on the boundaries of Weyl semimetals ~\cite{Arcs2015}. Such unusual phases are interesting from a fundamental point of view but also in light of potential applications. For instance spin-momentum locking may have uses in the field of spintronics ~\cite{Ilan2014,Wu2011,Ojeda2012}. 

Angle resolved photo-emission spectroscopy (ARPES) provides much of the ``smoking gun" evidence ~\cite{RMP_TI} for topological insulators and semimetals by directly measuring the electronic dispersion on their boundaries. A serious limitation is that these boundaries must be prepared in a careful and controlled manner to avoid distorting the electronic states. Consequently one tries to cleave only along a few natural planes while others would be disordered on the atomic scale ~\cite{Rasche2015} and presumably too rough to do ARPES on. At least for TIs, it is clear that states remain delocalized and metallic on such disordered boundaries ~\cite{RMP_TI}. Still ARPES on such boundaries was so far limited to disorder which was effectively weak ~\cite{Roushan,NotLikeRoushan}. 

Although the cleaving problem is prevalent, one class of topological materials in which it is clear and pressing are weak TIs ~\cite{FuKaneMele2007,Ringel2012}. These phases can be thought of as stacks of 2D TIs and consequently not all their surfaces are metallic. For instance a surface parallel to the stacking surface will be gapped. Unfortunately, for the compound $Bi_{14} Rh_3 I_9$ which appears more and more likely to be a weak TI ~\cite{Rasche2013}, such gapped surfaces are the only natural ones to cleave ~\cite{Rasche2015} and so while edges of this material clearly support some in-gap states ~\cite{Rasche2015} verifying that these are two counter chiral edge modes has not been possible so far. 

In this work the usefulness ARPES data on strongly disordered boundaries of several topological insulators and semimetals is examined analytically and numerically. Considering the ensemble of ARPES spectra generated by different disorder realizations, it is shown that while the average ARPES spectra does not contain useful information correlations within this ensemble may contain delta-function sharp features in momentum space. This occurs whenever electron transport retains its ballistic character in the face of disorder. Such is the case for integer quantum Hall effects, Chern insulators, and as shown below also Weyl semimetals. Provided time reversal symmetry (TRS) is maintained, sharp correlations also appear in 2D TIs and 3D weak TIs. For weak TIs one needs to cleave the weak TI such that it supports one or more 1D metallic edges which are decoupled from the rest thereby preventing backscattering and rendering them ballistic. In some other cases, for instance a strong 3D TI, these correlations will show a power law singularity. I also argue that using current nano-ARPES techniques ~\cite{Avila2013} such disorder correlation effects may very well be within experimental reach. 


We begin by analyzing the following ARPES setup on the edge of a 2D TI aligned (on average) parallel to the $\hat{x}$ direction. A photon beam of energy $\nu$ is shone on the edge and emits, via the photo-electric effect, electrons out of the material. Within the simple sudden approximation the detection rate $\Gamma(E,\vec{p},s)$ for an photo-electrons of $3D$ momentum $\vec{p}$, energy $E+\nu$, and spin $s$ probes the density of occupied states ~\cite{Damascelli2003} which in Green's function formalism is given by $f_D(E)\Im G_{ss}^r(E;\vec{k})$, where $\Im$ denotes the imaginary part, $G^r_{ss}$ is the retarded Green's function at crystalline momentum $\vec{k}$ corresponding to $\vec{p}$, and $f_D(E)$ is the Fermi-Dirac function. 

Introducing disorder on the edge one must verify that the sudden approximation still holds. The natural concern is lack of (crystalline) momentum conservation of the photo-electron due to disorder induced scattering. There are several complementary ways by which this issue may be controlled: First one may use energies ($\nu$) such that the photo-electron's wavelength is much smaller than the dimension of detects thereby reducing the amount of scattering. Assuming detects to be a few angstroms long with a strength ($V_0$) of a few $eV$ implies that for $\nu=400eV$ (corresponding to $0.6 \AA$) scattering effects should be strongly suppressed. Alternatively one can consider lower energies and limit the disorder to be cleaving-induced such that it occurs only on a length ($a$) of a few angstroms from the edge. A quantitative analysis of photo-electron scattering using the Born series is given in the Supp. Mat. A. Notably even in cases when these conditions are not strictly met it is likely that only the un-scattered component of the photo-electron would contribute in a coherent and sharp manner ~\cite{Durham1981}.

Given that photo-electron scattering effects are indeed suppressed and repeating the standard derivations ~\cite{Damascelli2003,Almbladh1985} one obtains  
\begin{align}
\label{Eq:ARPES}
\Gamma(E,\vec{p},s) &\approx f_D(E)\Im G_{ss}^r(E;\vec{k},\vec{k}).
\end{align} 
where ARPES now measures the {\it diagonal elements} in momentum space of the Green's function in the disordered system. 

The continuum theory for a 2D TI edge with TRS respecting disorder is given by ~\cite{TIEdge2006}
\begin{align}
H &= v_f i \partial_x \sigma_z+ V(x) 
\end{align}
where $\sigma_z$ is a Pauli-matrix in spin space, $V(x)$ is the disorder potential modeled here as Gaussian with $\langle V(x) V(y) \rangle = l^{-1} v_f^2 \delta(x-y)$ with $l$, the phase coherence length ~\cite{akkermans2007}, being a few atoms for strong disorder. The Green's function obtained from this theory is related to $G_{ss}^r(E;\vec{k},\vec{k})$ is 
\begin{align}
G_{ss}^r(E;\vec{k}) &\approx G_s^r(E;k_x),
\end{align}
which describes edge modes with well defined spin which are delta functions in the $y$-coordinates ($G_s^r(E;k)$ is short for $G_s^r(E;k,k)$). While this is obviously an approximation it would be shown to capture the essential physics. 

Conveniently the above disorder can be removed using a local gauge transformation ($\exp( \sigma_z \frac{i \int^x V}{v_f})$) and the eigenstates and energies are given explicitly by 
\begin{align}
\psi_{n,s} &= |s\rangle \times  
\frac{1}{\sqrt{L}} e^{\frac{i s \int^x V(x)}{v_f} + i k_n x} \\ \nonumber
E_{n_k,s} &= s v_f k_n 
\end{align}  
where, $s = \pm 1$ denotes the spin orientation in the $z$-direction and, without loss of generality, I assumed that the edge is compact and of length $L$ and thus supports discrete momentum eigenvalues $k_n = \frac{2 \pi n}{L}$. For simplicity I also assume that $\int V(x) = 0$ ~\cite{Doesnotmatter}. 

The resulting disorder averaged Green's function is then 
\begin{align}
&\Im \langle G^r_s(E;k) \rangle_{V} = \sum_n \delta (E- s v_f k_n) \int \frac{dx dy}{L^2} e^{ i \delta k_n  (x-y) } \times  \\ \nonumber 
&\left\langle e^{\frac{is \int_x^y V}{v_f}}\right\rangle_{V} = \sum_n \delta (E- s v_f k_n)  \int dx dy e^{ i \delta k_n (x-y) }  e^{-\frac{|x-y|}{2 l}}  
\end{align}
where $\delta k_n = k - k_n$, $\langle ... \rangle_V$ denotes disorder averaging, and the last equality is valid for $L \gg |x-y|$ and follows from standard manipulations of Gaussian integrals. Considering the limit $L \gg l$ allows us to trade $\int_{0}^{L} dx \int_0^{L} dy$ with $2 \int_0^{L} d w \int_0^{L} d W$, (where $w = (x-y)/2$ and $W=(x+y)/2$) while neglecting the dependence of the region of the $dW$-integration on $w$. This yields  
\begin{align}
\label{Eq:ImG}
&\Im \langle G^r_s(E;k) \rangle_{V} &= \sum_n \delta (E- s v_f k_n)  \frac{4 L^{-1} l^{-1}}{4\delta k_n^2+l^{-2}},
\end{align}
and exhibits no sharp signatures in momentum space. 

Looking for sharp signatures we turn our attention to the following type of correlation 
\begin{align}
C(E,E';k,k') &\equiv  \sum_{ss'} \langle \Im G_s(E;k) \Im G_{s'} (E';k') \rangle_{V} \\ \nonumber &- \langle \Im G_s(E;k)  \rangle_{V} \langle \Im G_{s'} (E';k') \rangle_{V}.
\end{align} 
Consider the 4 wavefunction average $A_4 = \langle \psi^*_{n,s}(k) \psi_{n,s}(k)\psi^*_{n',s'}(k') \psi_{n',s'}(k')\rangle$, appearing in the first term on the above r.h.s. 
\begin{align}
A_4 &= 4 \int_0^L \frac{ dwdw'dWdW'}{L^4} e^{i (w+w')(\delta k_n+\delta k'_n) + (w-w') (\delta k_n- \delta k_n')}  \\ \nonumber 
& \times \left \langle e^{i v_f^{-1} \left( \int_{W-w}^{W+w} V + \int_{W'-w'}^{W'+w'} V \right) } \right \rangle_V.   
\end{align} 

To analyze the above integrals, let us split their integration region into two sub-regions ($A,B$). The first would be that in which the integrals $[W-w,W+w]$ (or $[W+w,W-w]$ if $w<0$) and $[W'-w',W'+w']$ do not overlap and the second would be the complementary region. Within the first region, the random variables $ \int_{W-w}^{W+w} V$ and $ \int_{W'-w'}^{W'+w'} V$ are independent. Consequently it would not contribute to any correlation. Focusing on $B$, we further divide it $B_+$ and $B_-$, according to the sign of $ww'$. For $ww' > 0$ the two integral terms over the random variable $V$ add up instead of canceling. Consequently $|w|$ and $|w'|$ would be effectively limited to $|w|,|w'| <\approx l$ resulting in a minor and smooth $k-$space signature. 

Within the remaining $B_-$ region one obtains  
\begin{align}
[A_4]_{B_-} &=  \frac{1}{L^4} \int_{B_-} d\Sigma_+ d \Sigma_- d \sigma_+ d \sigma_- \\ \nonumber 
&\times e^{i(\delta k_n - \delta k_n') \sigma_- + i(\delta k_n + \delta k_n') \sigma_+}e^{-\frac{|\Sigma_-| + |\sigma_+|}{l}},
\end{align}
where $\Sigma_{\pm} = W \pm W'$ and $\sigma_{\pm} = w \pm w'$. The crucial point is that the $\sigma_-$ integration is not accompanied by any exponential damping factor and furthermore no such damping factors enter via the boundary dependence of the remaining integrals. The resulting singular part (in $k-$space) comes out to be (see Supp. Mat. B.)
\begin{align}
\label{Eq:A4}
A_4 &= \frac{1}{L^2}\frac{16}{(\delta k + \delta k')^2 + l^{-2}} \delta_{\delta k_n,\delta k'_n} + \{Regular\},
\end{align}
with $\{Regular\}$ denoting smooth functions of $k$. Comparing with $\Im \langle G_s(E;k) \rangle_{V}^2$ one finds that when $|\delta k|$ is smaller than $l^{-1}$, they are of the same magnitude and in particular their ratio does not scale with $L$. 

The above modeling is justified at the level an effective low energy theory which is expected to be universal. However when $k-k'$ becomes comparable to the size of the Brillouin zone this description may very well miss some features. To test this, $C(E,E';k,k')$ and $\Im G^r_{ss}(E;k)$ were evaluated for the Kane and Mele model ~\cite{KaneMele2005} in a cylindrical geometry having two zig-zag edges, a circumference of $320$ sites and a length of 14 sites. Parameters were taken to be generic with no extra symmetries ($t=1,\Lambda_{so}=1,\Lambda_{Rashba}=0.15,\Delta=0.2$) and bulk gap was $\approx 2t$. Potential disorder, uniformly distributed between $[-8,8]$, was placed on the first horizontal rows of sites near each edge. Its amplitude was tested to be the strongest possible in the sense that higher disorder actually pushes electrons deeper into the sample making the effective disorder weaker ~\cite{Ringel2012}. For each instance of disorder, an integral of $\Im G_{ss'}(E;y,y';k,k)$ over a $\Delta E$ energy window was numerically obtained. This energy window was chosen to have on average one state per momentum point. The spinless $1D$ density of states ($\Im G^r(E;k) = \sum_s \Im G^r_{s}(E;k)$) was obtained by tracing over spin indices and half of the $y$ indices ($\sum^{y=5}_{s,y=0} G^r_{ss}(E;y,y;k,k)$) thereby incorporating states from only one edge of the cylinder. Note that several other tracing method were tested as well as different disorder profiles and longer cylinder lengths. These had no qualitative effect on the results shown below. For statistics, $4000$ disorder realizations were used.

\begin{figure}[ht!]
\includegraphics[width=\columnwidth,clip=true,trim= 20 20 500 20]{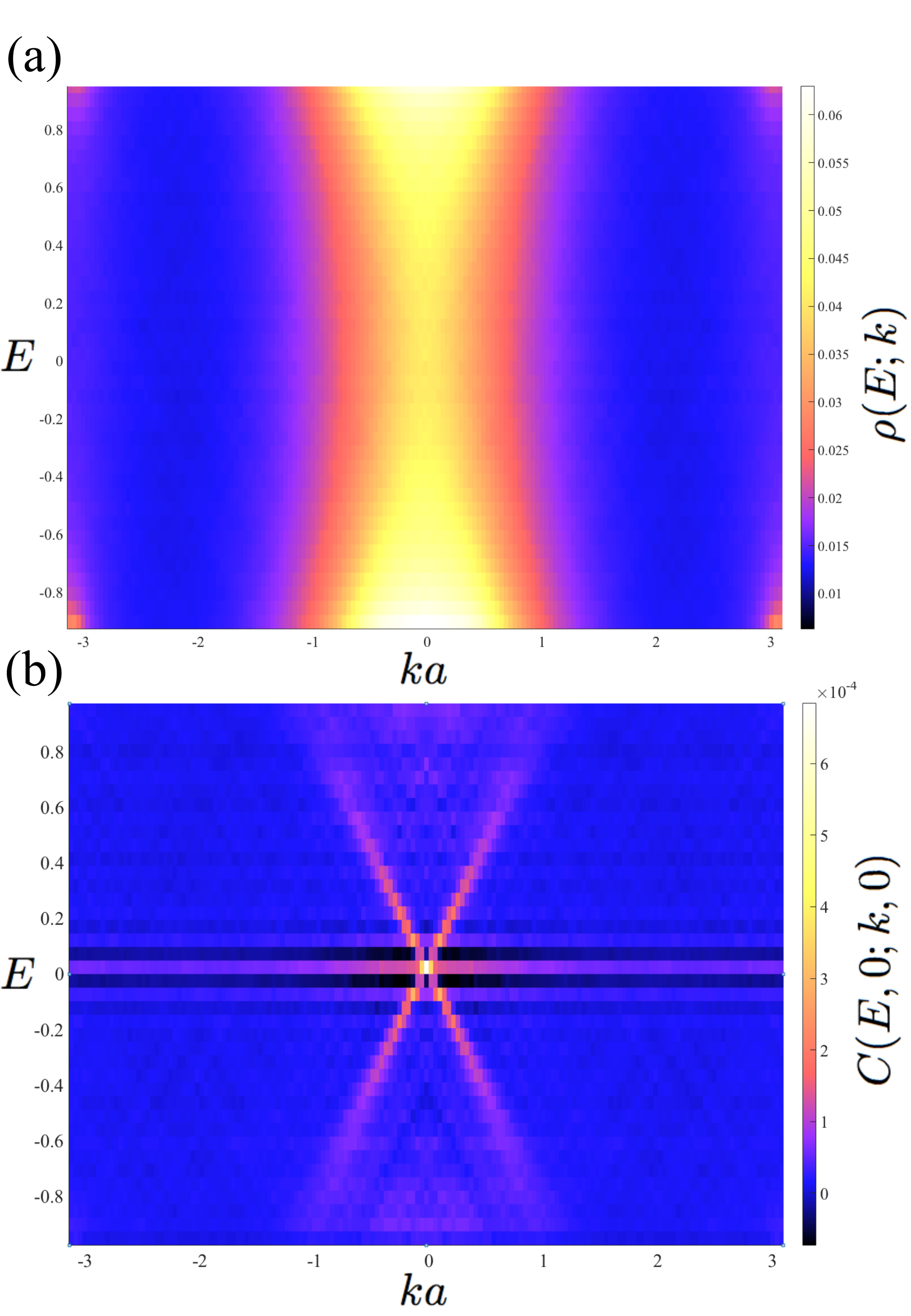}
\caption{(a) Numerically obtained density of states ($\rho(E;k)$) as a function of momentum ($ka$) and energy ($E$) on an edge of a topological insulator with strong disorder. The energy scale here is the bulk gap. This quantity shows no sharp features. (b) For the same data the correlation $C(0,E;0,k_n)$ between $\rho(E;k)$ and $\rho(0;0)$ does show delta-function sharp signatures reflecting the ballistic nature of electron transport on a topological insulator edge (statistical error is roughly $1 \cdot 10^{-5}$). \label{Fig:Numerics}}
\end{figure}

The numerically obtained Green's function averages and correlation functions are shown in Fig. (\ref{Fig:Numerics}). The upper plot shows the $\rho(E;k) = \langle \Im G^r(E;k_n) \rangle_V$, as a function of $n$ (wavenumber) and $E$ (in units of $t$).  The underlying $1D$ Dirac spectrum (see Supp. Mat. Fig. 3.) has become completely blurred. The lower plot shows $C(0,E;0,k_n)$ for the same system where a Dirac like dispersion clearly reappears.  It was numerically verified these that features become delta-sharp as $L \rightarrow \infty$ for small $E$ and $k_n$ in consistency with the previous argument about $P(\omega,q)$. An additional feature which is discussed in Supp. Mat. Sec. C is a horizontal line around $E=0$ which appears also for a band insulator (see Supp. Mat. Fig. 4) and reflects the presence of by-standing localized states  

Lastly a direct link can be establish between such sharp correlations and ballistic motion of particles. Indeed rewriting $C(E,E';k,k')$ as $C(E,E+\hbar \omega;k,k+q)$ and integrating over $E$ one obtains   
$\int dE \langle G^r(E;k) G^a(E+\hbar \omega,k+q)\rangle_V$ plus its complex conjugate [the $G^r G^r$ ($G^a G^a$) terms vanish by closing the energy contours on the upper (lower) half complex plane]. On the other hand the {\it probability of diffusion} ~\cite{akkermans2007} is given by $P(\omega,q)=\int dE dkdk' \langle G^r(E;k,k')G^a(E+\hbar\omega;k+q,k'+q)\rangle_V$. At strong disorder ergodicity of momentum ensures that the $k$ and $k'$ integrations are superfluous ~\cite{akkermans2007}. As a result  
\begin{align}
\int dE C(E,E+\hbar \omega;k,k+q) &\propto \Re P(\omega,q) +\{Regular\},
\end{align} 
where $\Re$ denotes taking the real part. For a regular conducting system and near $\omega,q \rightarrow 0$, $P(\omega,q) = -i\omega + D q^2$ ~\cite{akkermans2007} and no delta sharp features are obtained, only power law singularities. In contrast on a TI edge one expects a ballistic behavior and so $P(\omega,q)$ should behave as $\sum_s (i\omega-i s v_f q - \epsilon)^{-1}$ which indeed yields a sharp delta function. Diffusion of such waves around their center of motion can be included by adding a $D q^2$ term to the denominator which would be irrelevant at sufficiently low momentum. Such a term should appear for a dispersive spectrum therefore did not show up in Eq. (\ref{Eq:A4}). 

%


Generalizing the above for IQHE and Chern insulators is trivial and amounts to ignoring the spin index. Considering a $3D$ weak TI the above results remain valid in the following sense. The bulk $3D$ weak TI can be thought of as coupled layers of $2D$ TIs. Consequently if one carves out just the upper layer (or any finite odd number of layers) an effective $2D$ TI edge is created and the previous results apply. Considering Weyl semimetals, their universal feature is surface Fermi arcs which are inherently chiral. Transport on these arcs is limited to directions perpendicular to the arc. Nonetheless electrons can scatter away from the arc either into the gapless bulk or perhaps onto some accidental boundary modes. Since the bulk is assumed to be a semimetal, the density of states near the band touching energy should be strongly reduced, ideally to zero. If one is unfortunate enough to have accidental boundary modes near this energy, it is still possible these would localize and become effectively removed. Consequently sharp correlation features should appear in generic Weyl semimetals.     

\begin{figure}[ht!]
\includegraphics[width=\columnwidth,clip=true,trim=10 30 535 20]{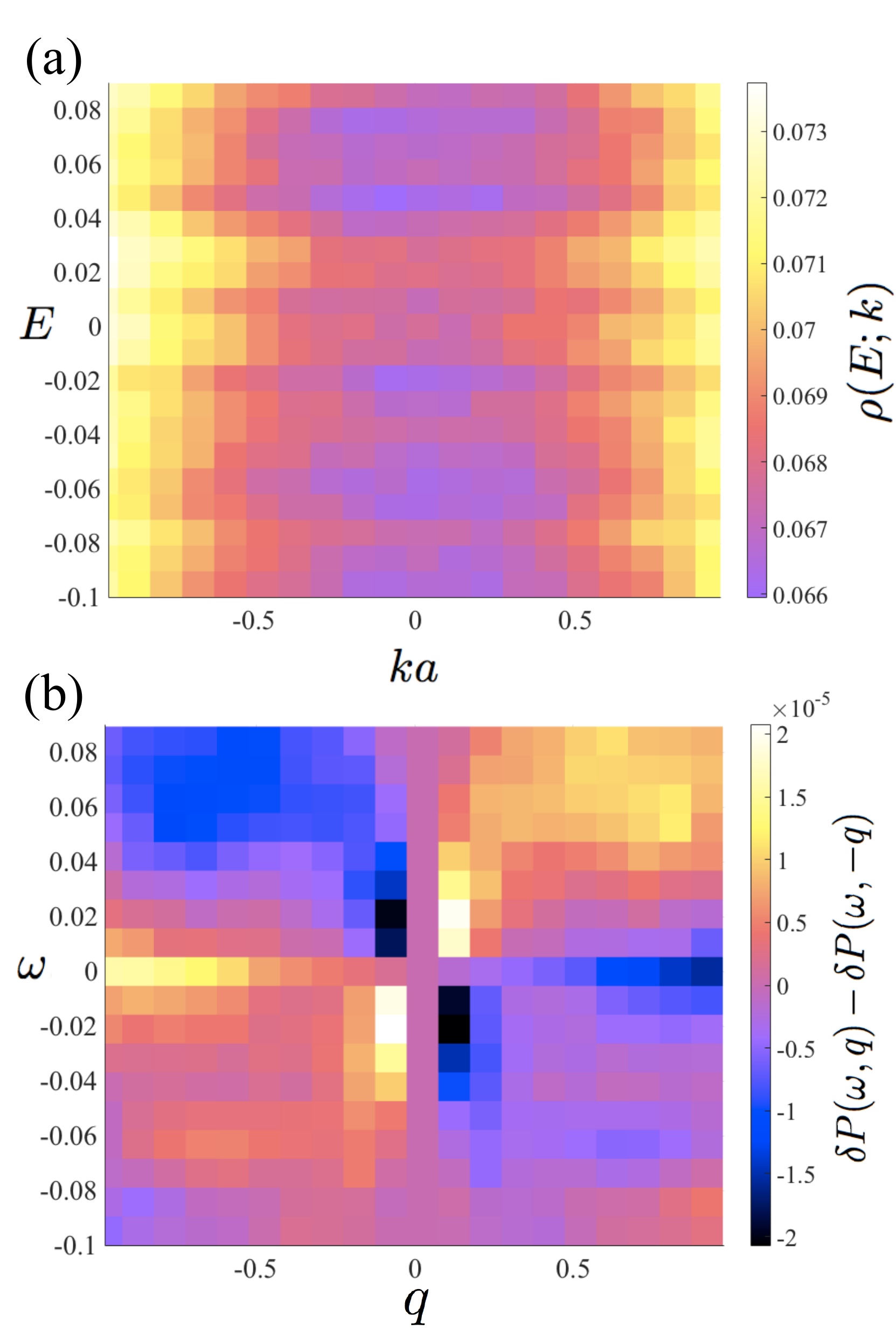}
\caption{(a) Numerically obtained density of states on a disordered, $60\times60$ unit-cells surface of a Weyl-semimetal as a function of momentum ($ka$) in the direction ($\hat{k_1}$) perpendicular to the Fermi arc and integrated over the orthogonal momenta. (b) The asymmetric piece of the ARPES correlation $\int dE dk C(E,E+\omega;\vec{k},\vec{k}+q \hat{k_1})$. Correlations again show a sharp features on the scale of a few wave numbers. Note that only $1/3$ of the Brillouin zone is displayed. Statistical error is roughly $4 \cdot 10^{-6}$.  
\label{Fig:Numerics2}}
\end{figure}

As detailed in Supp. Mat. Sec. D. the quantity $\delta P(\omega,q)=\int d^2 k dE C(E,E+\hbar \omega;\vec{k},\vec{k}+q \hat{k_1})$ was numerically obtained on a surface of a particular Weyl semimetal with $\hat{k_1}$ being roughly perpendicular to its Fermi arc. Its asymmetric components $\delta P(\omega,q)-\delta P(\omega,-q)$ and the density of states are shown in Fig. (\ref{Fig:Numerics2}). Notably the average is again featureless while correlations have sharp features. Furthermore calculating $\delta P(\omega,q)-\delta P(\omega,-q)$ with $q$ parallel to the Fermi arc or for a semimetal without Fermi arcs such sharp features are absent (see Supp. Mat. Fig. 7.).  

Several comments are in order concerning contact with experiment. Typical bulk-gap energy scales for TIs range between a few $meV$ ~\cite{konig2007} and a few hundred $meV$ ~\cite{Rasche2013}. The ratio of band width scales ($\approx 1eV$) to these gap scales indicates an effective boundary modes depth of roughly $1-100nm$. A nano-ARPES beam has finite spot profile $B(r)$ around $100nm$ ~\cite{Avila2013}. In effect this simply replaces the $L$ used above with an effective $L_{eff}$ equal to the size of the beam. Experiments also measure $\Im G^r(E;\vec{k})$ integrated over a small momentum window ($\Delta k$) due to finite resolution. Apart from reduction in resolution, this reduces $C(E,E';k,k')$ by a factor of $N^{-1}$, where $N=\Delta k L_{eff}$ counts the number of different $k_n$ points being averaged together in each measurement (considering realistic values of $\Delta k=0.005A^{-1}$ ~\cite{Avila2013} gives $N=5$). Similar issue arises given any form of averaging {\it independent subsystems} within one measurement, caused for instance by measuring several decoupled edges together. Counter-intuitively increasing $N$ has no damaging effect on the relative statistical error when obtaining $C(E,E';k,k')$ using variance estimators \cite{VarEstimators}. 
Also integrating $C(k,k+q,E,E+\omega)$ over $k,E$ thereby obtaining $\Re P(\omega,q)$ greatly reduces the number of samples needed for a good variance estimation. For example, calculating $\Re P(\omega,q)$ for the 2D TI yielded clear features even for just a single disorder realization. For the Weyl semimetal $45$ disorder realization were used to generate Fig. (\ref{Fig:Numerics2}). The main concern of having $N \gg 1$ is the inability to distinguish the variance related to disorder ($\sqrt{C(E,E';k,k')}$) from extrinsic sources of noise. Notably for the $2D$ TI considered here the ratio of $\rho$ to $\sqrt{C(E,E';k,k'}$ was $1:1$ as one can verify by comparing these quantities in Fig. (\ref{Fig:Numerics}). For the Weyl semimetal this ratio was roughly $1:25$. Removing the $k$ independent contribution from $\rho$, coming most likely from localized states, this ratio was roughly $1:5$.  

To conclude while $\rho(E;k)$ averaged over disorder realizations is featureless in momentum space, the variance of $\rho(E,k) \rho(E',k')$ can have delta-function sharp features in momentum space. Such features appear whenever propagation of density waves has ballistic character which is the case for various topological phases even when disorder is strong. Measuring these correlations may improve the accuracy of ARPES and allow its implementation in cases where clean cleaving is difficult to achieve. 

\vspace{5mm}
\textbf{Acknowledgments} The author would like to thank Niels Schroeter, Thomas Scaffidi, John Chalker, Steven H. Simon and Fabio Caruso for helpful comments and discussions. This work was supported by the European Union's Horizon 2020 research and innovation programme under the Marie Sklodowska-Curie grant agreement No. 657111.


\begin{thebibliography}{10}
\bibitem{NotLikeRoushan}
In the previous Ref. disorder was not strong enough to blur the average ARPES
  data and the equal energy correlations which were considered. See Supp. Mat.
  for a detailed comparison..
\bibitem{Doesnotmatter}
Alternatively, redefine $V(x) \rightarrow V(x) - L^{-1} \int dx' V(x');E_{k,s}
  \rightarrow E_{k,s} + L^{-1} \int dx' V(x')$. The extra energy fluctuation
  can be treated straightforwardly in all which follows.
\bibitem{VarEstimators}
Given samples $x_{1..M}$, variance ($\sigma^2$) is estimated using $s =
  \sum^M_n ( x_n - \bar{x})^2/(M-1)$, with $\bar{x} = \sum^M_n x_n / M$.
  $Var[s^2]$ is then $2\sigma^4/M-1$ and does not depend on the average.
\end{thebibliography}



\section{Supplemental Material}

\subsection{Suppression of photo-electron scattering for disorder confined to the edge}
As discussed in detail in Ref. (\onlinecite{Almbladh1985}) what one measures in ARPES is rate of emission of time reversed LEED states ($|\chi^{-}_p\rangle$). These states appear as plane waves only asymptotically. Closer to the sample scattering from the crystal and, in our case, disorder augment their form. This can be analyze through the Lippmann-Schwinger equation 
\begin{align}
\langle r |\chi_p^-\rangle &= \frac{1}{(2\pi)^{3/2}} e^{i \vec{p} r} - \frac{2m}{\hbar^2} \int d^3 r' \frac{e^{ik_E|r-r'|}}{4\pi |r-r'|} V(r') \langle r'|\chi_p^{-}\rangle 
\end{align}
where $V$ is the potential induces by the crystal and the disorder. Focusing here only on the effects of disorder, $V$ is taken to be 
\begin{align}
V(r) &= a^2 \delta(r_z) \delta(r_y) \tilde{V}(r_x) 
\end{align}  
where $a$ is the dimension of the edge in the $x$ and $y$ directions and $V(r_x)$ is some random function with $\langle \tilde{V}(x)\tilde{V}(x') \rangle = V_0^2 g(x-x')$. To first order in perturbation theory one can replace $|\chi_p^{-}\rangle$ by $\frac{1}{(2\pi)^{3/2}} e^{i \vec{p} r}$ and perform the $r_x$ and $r_y$ integrals to obtain 
\begin{align}
\label{Eq:LS}
\langle r |\chi_p^-\rangle &= \frac{1}{(2\pi)^{3/2}} \left( e^{i \vec{p} r} - \frac{2m a^2}{4\pi\hbar^2 } \int d x' \frac{e^{ik_E|r-x'|}}{|r-x'|} \tilde{V}(x') e^{i \vec{p} r}\right) 
\end{align}
Considering $\vec{p}=(0,p_y,p_z)$ and examining the disordered region ($y=z=0$) one obtains 
\begin{align}
\langle x |\chi_p^-\rangle &= \frac{1}{(2\pi)^{3/2}} \left( e^{i \vec{p} r} - \frac{2m a^2}{4\pi\hbar^2 } \int d x' \frac{e^{ik_E|x-x'|}}{|x-x'|} \tilde{V}(x')\right).
\end{align}
The integral is a random number whose variance at each point ($\sigma^2$) is given by 
\begin{align}
\sigma^2 &= \int dt ds \frac{e^{ik_E|t|-ik_E|s|}}{|t s|} \langle \tilde{V}(t) \tilde{V}(s)\rangle  \\ \nonumber 
&=V_0^2 \int dt ds \frac{e^{ik_E|t|-ik_E|s|}}{|t s|} g(t-s)
\end{align}
For short range correlated disorder $g(x-y)$ falls exponentially after some characteristic distance $l \approx a$. Consequently the integral converges for large $t$ and $s$. For small $t,s$ its logarithmic divergence is cut-off by the size of the edge $a$. Consistently with our assumption of a delta function transverse disorder profile we take $k^{-1}_E \gg a$ and obtain 
\begin{align}
\sigma^2 &= 2 \pi V_0^2 \log(l/a) \approx 2 \pi V_0^2 
\end{align}
Re-examining Eq. (\ref{Eq:LS}) one finds that the ratio of the zero and first order terms is $\epsilon=\frac{2m V_0 a^2}{\sqrt{8 \pi} \hbar^2}$. Taking realistic values of $a=5\AA$ and $V_0=5eV$ gives $\epsilon \approx 1/5$. Perturbation theory, or the Born series, is then expected to converge rapidly and the Fourier transform of $|\chi_p^{-}\rangle$ would have a dominant sharp pick at momentum $p$. Notably at $100eV$ the de Broglie wavelength is $1.2\AA$ and becomes shorter than $a$. This should further improve the convergence of the Born series.

\subsection{Integration over the $B_-$ region}
The $B_-$ region is defined by $ww'<0$, $W,W' \in [0,L]$, $w,w' \in [-L/2,L/2]$, and $[W-w,W+w]\cap[W'-w',W'+w']\neq \emptyset$. It may be further split to according to the sign of $w$ and the sign of $W-W'$. Since the integral is symmetric with respect to latter sign, I can choose $W-W'>0$ and add a factor of $2$. The effect of the former is simply to flip the sign of $(\delta k_n - \delta k_n')$. Accordingly the integral domain for $w>0$ is   
\begin{align}
&2 \int_0^{2L} \Sigma_+ \int_0^{L/2} \sigma_- \int_0^{\sigma_-} \Sigma_- \int_{-\sigma_-}^{\sigma_-}\sigma_+ \\ \nonumber 
&+2 \int_0^{2L} \Sigma_+ \int_{L/2}^{L} \sigma_- \int_0^{\sigma_-} \Sigma_- \int_{\sigma_- - L}^{L- \sigma_-}\sigma_+
\end{align}
while as we recall the integrand is 
\begin{align}
e^{i(\delta k_n - \delta k_n') \sigma_- + i(\delta k_n + \delta k_n') \sigma_+}e^{-\frac{|\Sigma_-| + |\sigma_+|}{l}}
\end{align}
Due to the rapid decay of the integrand as a function of $\sigma_+$ and $\Sigma_-$, up to $O(l/L)$ corrections one can approximate the boundaries of integration as
\begin{align}
&2 \int_0^{2L} \Sigma_+ \int_0^{L/2} \sigma_- \int_0^{\infty} \Sigma_- \int_{-\infty}^{\infty}\sigma_+ \\ \nonumber 
&+2 \int_0^{2L} \Sigma_+ \int_{L/2}^{L} \sigma_- \int_0^{\infty} \Sigma_- \int_{-\infty}^{\infty}\sigma_+ \\ \nonumber 
&= 2 \int_0^{2L} \Sigma_+ \int_{0}^{L} \sigma_- \int_0^{\infty} \Sigma_- \int_{-\infty}^{\infty}\sigma_+
\end{align}
Carrying out the rest of this computation yields the result appearing in the main text.

\subsection{Additional numerical results for a 2D topological insulator} 
\begin{figure}[ht!]
\includegraphics[width=\columnwidth,clip=true,trim=0 380 500 0]{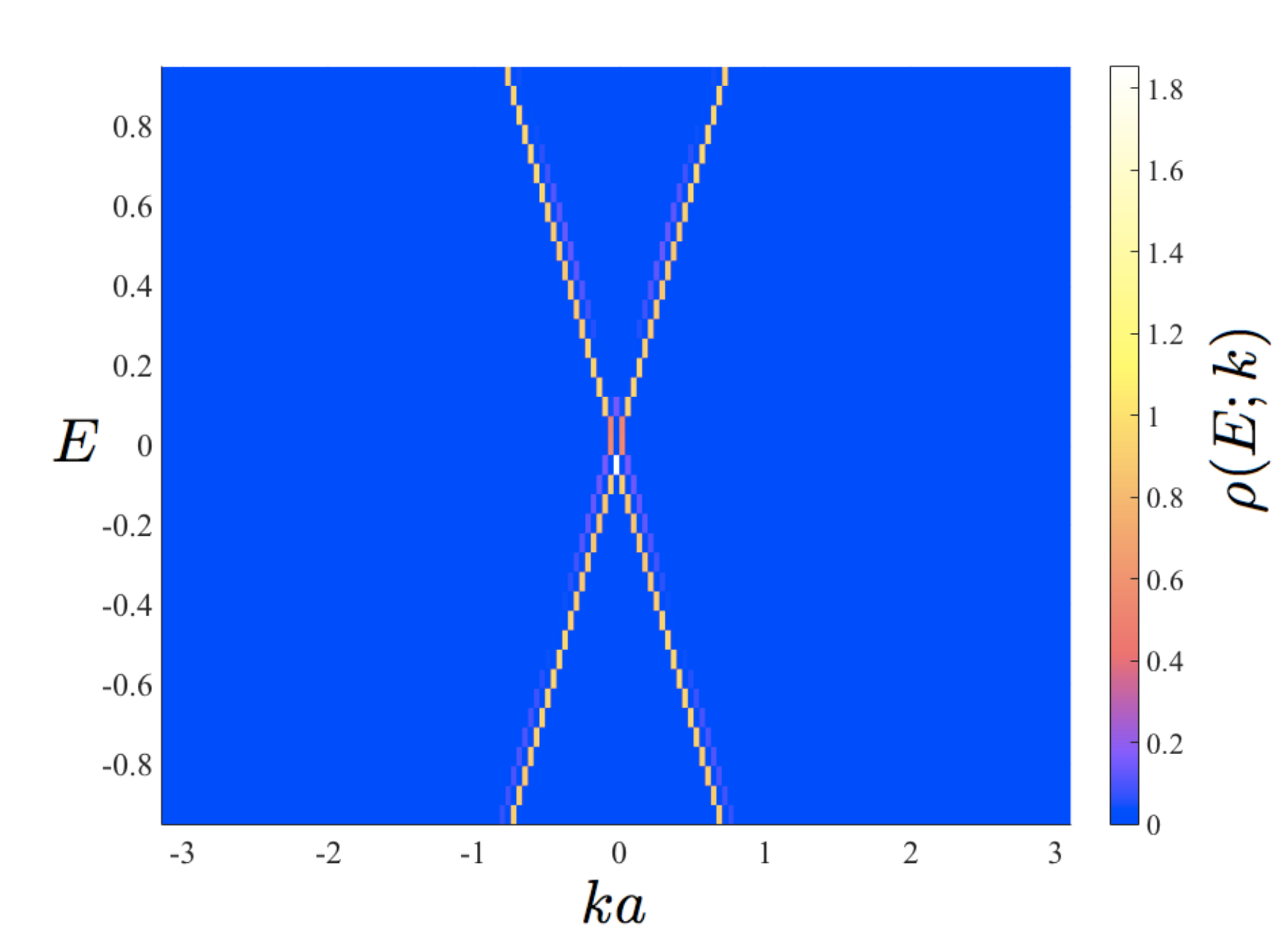}
\caption{The spectrum of the same system studied in the main text in the pristine case (no disorder). The velocity extracted from this spectrum is about $\%75$ of that obtained from ARPES correlations. A faint, order of magnitude weaker, signal from an extra Dirac cones appears due to tunneling from the opposite surface. Note that since the plots collects data from the half system close to one of the surface, this implies that coupling between the two Dirac cones is roughly suppressed by a $10^{-2}$. \label{Fig:Clean}}
\end{figure}

\begin{figure}[ht!]
\includegraphics[width=\columnwidth,clip=true,trim=0 380 500 0]{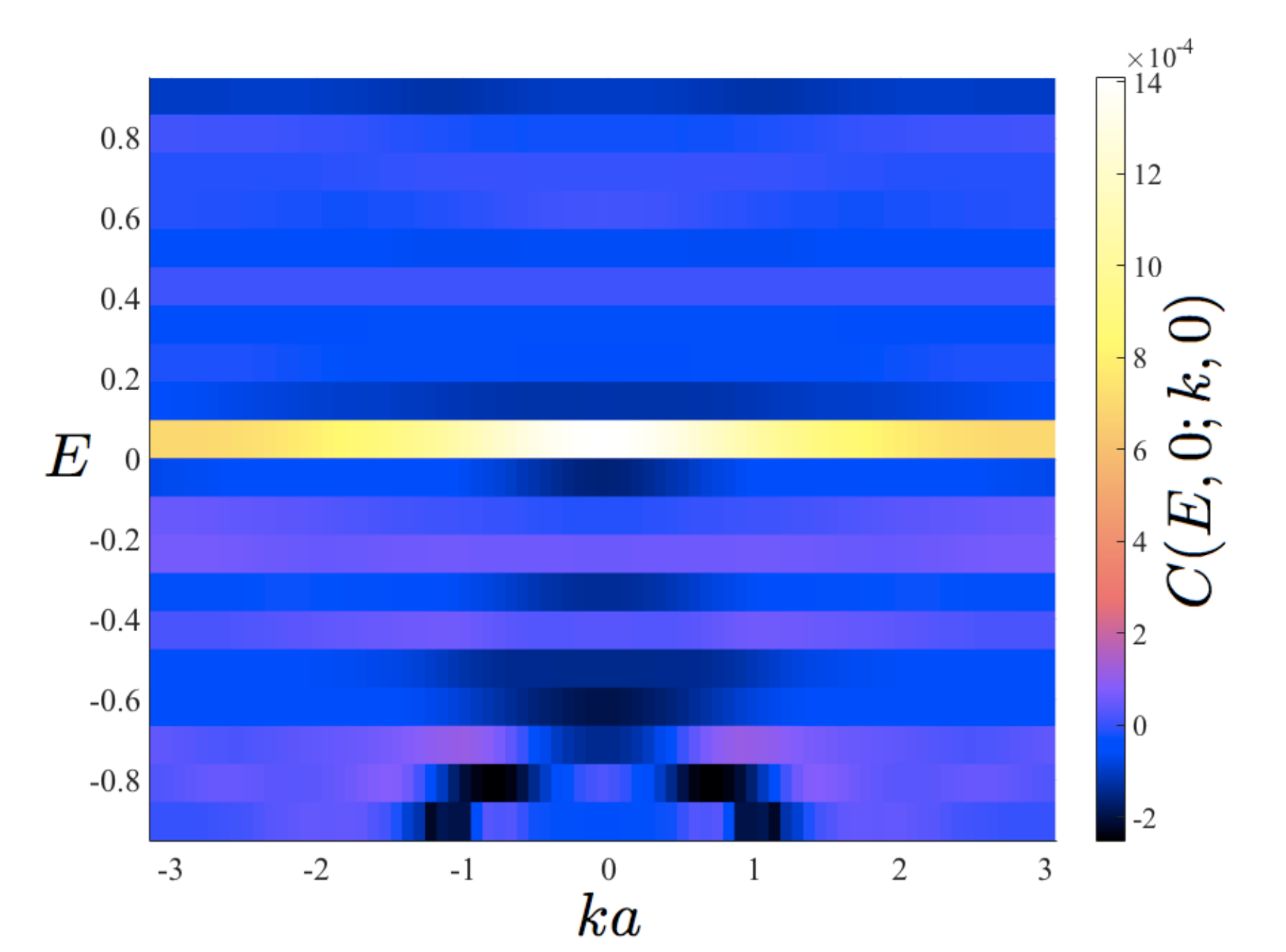}
\caption{ Numerically obtained density of states on the edge of a band insulator with strong disorder. Correlations do not exhibit any sharp feature in momentum space. \label{Fig:Band}}
\end{figure}

This section contains various additional numerical results. Figure (\ref{Fig:Clean}) shows the clean ARPES spectrum of the same system used in the main text. Figure (\ref{Fig:Band}) shows ARPES averages and correlations for the Kane and Mele model in the trivial phase ($t=1,\Delta=1,\lambda_{SO}=\lambda_R = 0$) with the same type of disorder considered in the main text. The plots are averages of 800 disorder realizations and the system size is $14$ by $160$ sites. 

The average ARPES spectrum again lacks any sharp feature and it is not particle hole symmetric due to the sub-lattice symmetry breaking induced by the zig-zag edge. Correlations of the ARPES spectra do contain sharp features but only as a function of $E$. This last feature, which occurred also in Fig. (\ref{Fig:Numerics}) of the main-text, can be attributed to localized states. Indeed consider $\Im G^r(E;k)$ for a particular disorder instance. Due to Anderson localization, the band insulator edge supports only localized states $\phi_n$ which may occur at random in-gap energies ($E_n$). Following this one can approximate $\Im G^r(E;k) \approx \sum_n \delta(E-E_n) \phi^*_n(k) \phi_n(k) \approx L^{-1} \sum_n \delta(E-E_n)$. Self correlations of an energy with itself in this expression yield a sharp $\delta(E-E')$ signature. Notably however, it appears that for energies just below and above $E=0$ there is some small degree of anti-correlation. This can be attributed to level repulsion coming from finite size effects. Indeed it was verified that this effect diminishes with increasing $L$.

\begin{figure}[ht!]
\includegraphics[width=\columnwidth,clip=true]{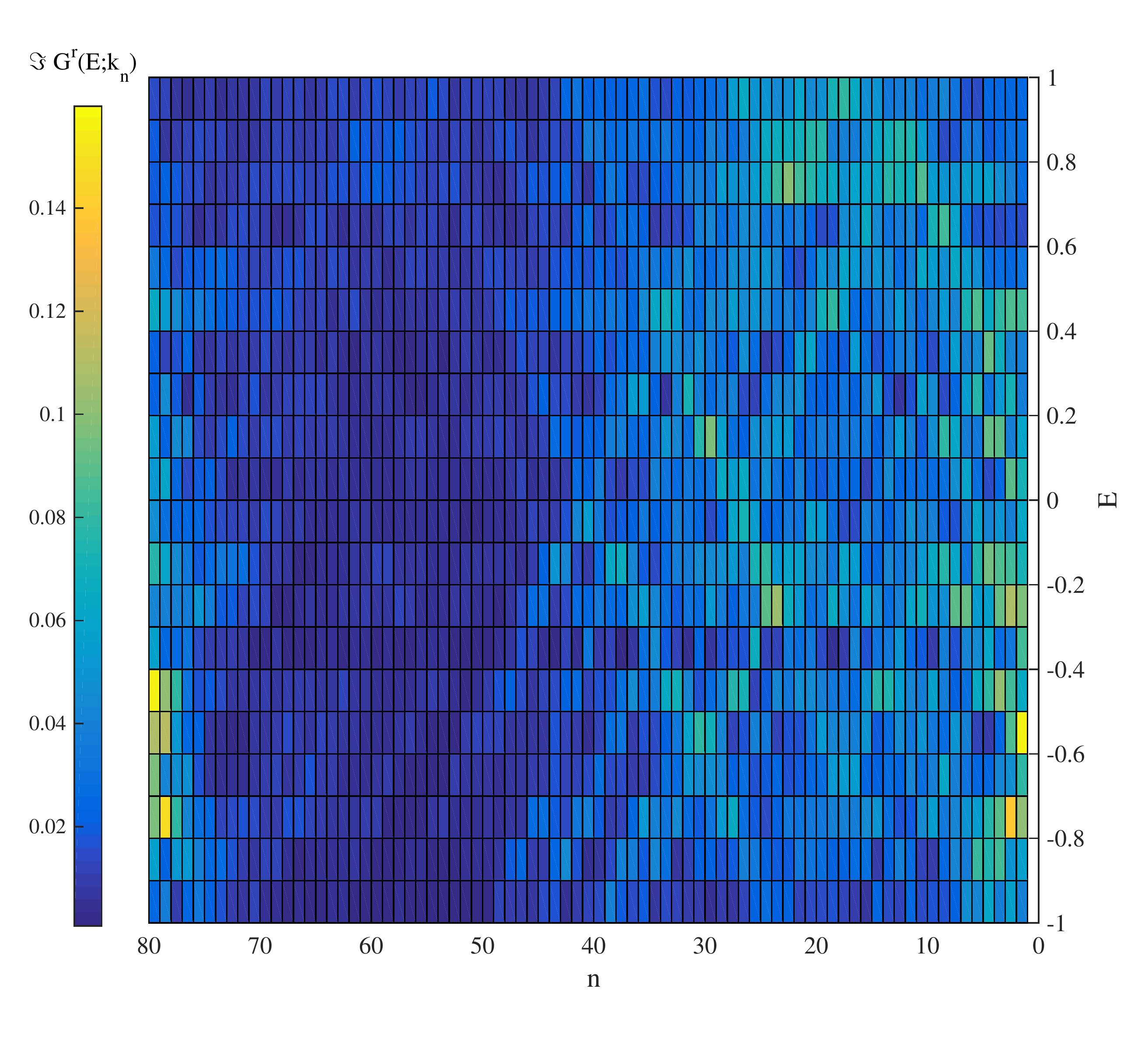}
\caption{Numerically obtained density of state of a smaller version ($10$ by $160$) of the same system studied in the main text for one randomly chosen disorder configuration. It appears as a noisy superposition of many Dirac cones. When averaged upon it becomes featureless however the correlation used in the main text exposes its hidden sharp nature. \label{Fig:One}}
\end{figure}

Lastly, Fig. (\ref{Fig:One}) shows $\Im G^r(E;k_n)$ calculated for one instance of disorder with the same parameters used in the main text except the system size which was $10$ by $160$. Evidently the data is very noisy and one might even guess that the Dirac cones is around $n=0$ rather than $n=40$. Notwithstanding the correlations in the data are already visible and are reflected by faint diagonal lines. Loosely speaking it is as if that the Dirac cone dispersion remains robust in shape but its location jumps between different regions. The correlation function used in the main text is designed to expose this type of correlation.    

\subsection{Numerical study of ARPES correlation in a Weyl semimetal} 
This section explains the model and methods used in obtaining the correlation of the density of state on the surface of a Weyl semimetal. As model for a semimetals with Fermi arcs, I use a slight generalization of the one introduced in Ref. (\onlinecite{FuKaneMele2007}). It is a tight-binding isotropic nearest neighbor hopping model on a diamond lattice with additional next nearest neighbor spin orbit and a Zeeman field 
\begin{align}
H_{W} &= t \sum_{\langle ij \rangle} c^{\dagger}_i c_j + i \lambda_{SO} \sum_{\langle \langle i j \rangle \rangle} c^{\dagger}_i \vec{s} (d^1_{ij} \times d^2_{ij}) c_j + B_z \sum_i c^{\dagger}_i s_z c_i
\end{align}
where $d^{1,2}_{ij}$ are the bond vectors connecting the next nearest neighbor sites $i$ and $j$, and $\vec{s}=(s_x,s_y,s_z)$ are Pauli matrices acting in spin space. As shown in Ref. (\onlinecite{FuKaneMele2007}) this model, at $B_z=0$, supports three Dirac cones at the three $X^{x/y/z}$ points in the Brillouin zone. Using the low energy theory they derived one finds that $B_z \neq 0$ splits the Dirac cone at $X^z$ into two Weyl cones at zero energy ($E=0$). The other two Dirac cone develop a ring like band touching at zero energy. Due to this peculiarity, the density of state near $E=0$ vanishes as $E$ rather than $E^2$. Consequently this specific Weyl semimetal is not ideal for reducing bulk leakage effects but still reasonably good. The parameters used throughout the numerics were $t=1;\lambda_{SO}=1;B_z=1$. 

\begin{figure}[ht!]
\includegraphics[width=\columnwidth,clip=true]{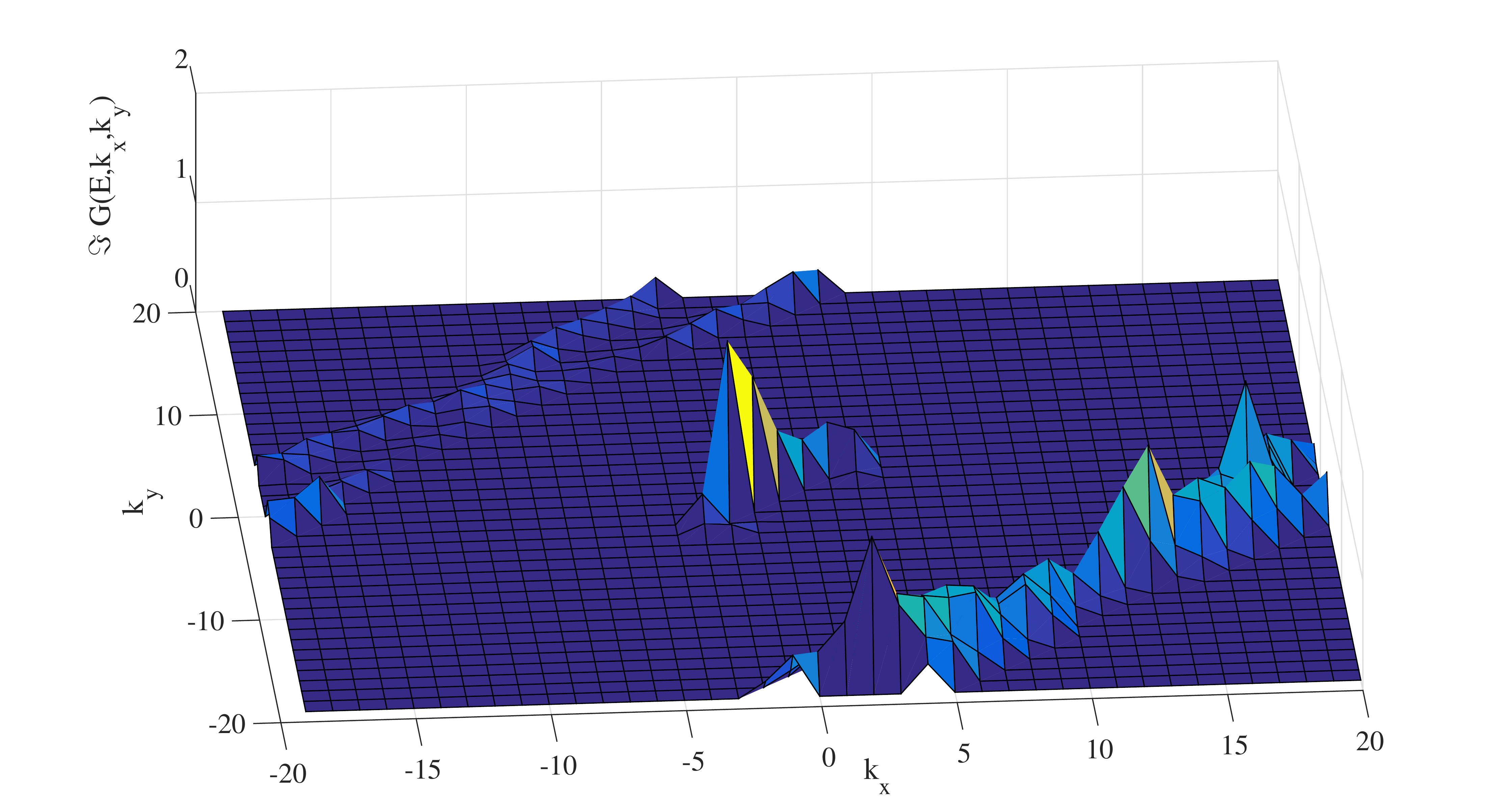}
\caption{Numerically obtained density of states on a pristine surface of the diamond based Weyl-semi metal slightly above the Dirac point ($E=0.1$) as a function of the two good momenta along the surface. The center of the plot shows the Fermi arc which, if one was to increase $E$, would gradually bend and travel in the direction $-
\hat{x}+\hat{y}$. The other features corresponds to the ring band touching developing around the other two Dirac codes. Scanning $E$ one can verify that they have no chiral features. 
\label{Fig:clean40}}
\end{figure}

Disordered surface state were generated as followed. First the three primitive lattice vectors of the diamond's FCC lattice were thought of as the $\hat{x},\hat{y},\hat{z}$ directions. The lattice was placed on a 60 by 60 units cells periodic geometry along the $\hat{x}$ and $\hat{y}$ axis, and with an 10 unit cells across open geometry on along $\hat{z}$. As shown in Fig. (\ref{Fig:clean40}), the resulting surface showed a Fermi arc with a Fermi velocity mainly in the $\hat{k_1} \propto -\hat{x}+\hat{y}$ direction. Strong potential disorder, uniformly distributed in the interval $[-8,8]$ was introduce on the first two layers at and just beneath each of the two surfaces. Calculation of $\Im G^r(E;k_x,k_y)$ was performed in the same way it was carried for a 2D TI however only the outmost layer near one of the surfaces was considered. The quantity $\delta P(\omega,q)$ was obtaining using 
\begin{align}
&\sum_{h,a,b} \Im G^r(\frac{h}{10}+\omega;k_{a+b+q},k_{a-b-q}) \Im G^r(\frac{h}{10};k_{a+b},k_{a-b}) \\ \nonumber & - \langle \Im G^r(\frac{h}{10}+\omega;k_{a+b+q},k_{a-b-q})\rangle \langle \Im G^r(\frac{h}{10};k_{a+b},k_{a-b})\rangle
\end{align}
with $h \in [-90,90],a \in [16,44],b \in [0,5]$, $k_n$ is the momentum at wavenumber $n \in [1..60]$ for the $60 \times 60$ surface. Reducing the range of $h$ to say $[-20,20]$ and $[-10,10]$ reduces the sharp pick to about $\%80$ and $\%60$ of its height. This is consistent with the expectation that only the modes close to the band touching point ($h=0$) contribute to ballistic transport. Other modes just leak too quickly into the bulk.

\begin{figure}[ht!]
\includegraphics[width=\columnwidth,clip=true,trim=0 10 535 10]{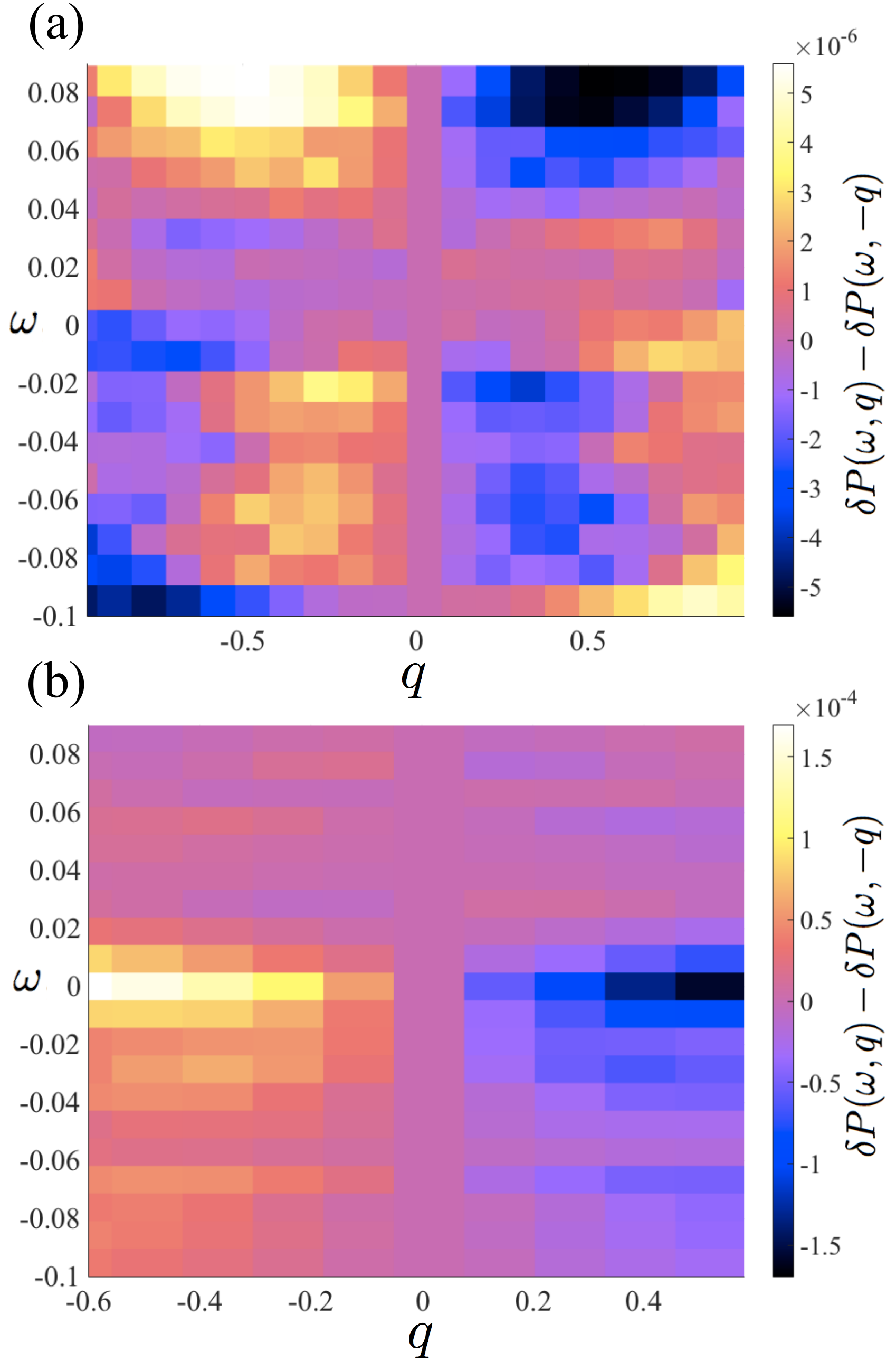}
\caption{The asymmetric part of the correlation function $\delta P(\omega,q) - \delta P(\omega,-q)$ for two cases where one does not expects ballistic behavior. (a) shows this quantity for the same system studied in the main text however with $q$ pointing parallel to the Fermi arc. Notably its an order of magnitude weaker compared to the correlations in the perpendicular directions shown in the main text. (b) The same quantity for a slightly smaller semimetal ($40\times40\times10$ unit cells) with $B_z=0$ and no Fermi arcs. In both cases no sharp features are observed in momentum space. 
\label{Fig:FalsePositives}}
\end{figure}

The upper part of Fig. (\ref{Fig:FalsePositives}) shows the quantity $\delta P(\omega,q_2)-\delta P(\omega,-q_2)$ which is the same in all aspects to that shown in main text except that the momentum direction here is parallel to the Fermi arc.  Consequently there is no reason to expect any ballistic features along this direct and indeed the correlations lack any sharp features (note the different in color map here compared to the main text). The lower part of Fig. (\ref{Fig:FalsePositives}) returns to study $\delta P(\omega,q_1)-\delta P(\omega,-q_1)$ however this time for a system with no Fermi arc ($B_z=0$) as well as for a slightly smaller system (40x40x10 unit-cells). Notably although the system preserves time reversal symmetry the surface lacks inversion symmetry ($(\hat{x},\hat{y},\hat{z}) \rightarrow (-\hat{x},-\hat{y},\hat{z})$) due to the spin-orbit term. The correlations just reveal a sharp feature in energy space, most probably coming again from localized states, but are otherwise smooth in momentum space. Data was obtained using 45 and 86 disorder instances for the two plots respectively.

\subsection{The Fourier transformed STM approach}

It is important to discuss some relations with the elegant approach used in the experiments of Ref. (\onlinecite{Roushan}). In that work the Fourier transform of real-space STM data ($JDOS(q)$) on $3D$ strong topological insulator surface with disorder was measured experimentally. This quantity showed sharp picks reflecting Friedel oscillations between strong scatterers. It was further compared with $\sum_k \Im G^r(E;k)T(E;q) \Im G^r(E;k+q)$ (for one instance of disorder, with $\Im G^r(E;k+q)$ obtained from ARPES) and found to agree well for some suitable choices of $T(E;q)$. Notably up to this $T(E,q)$ factor this quantity is the same as $P(\omega=0,q)$ studied in the main text. 

There are two qualitative difference between the measurement carried for that work and the ones suggested here. First, in Ref. (\onlinecite{Roushan}) the ARPES data itself already contains features comparable in sharpness to $JDOS(q)$. In contrast here the ARPES spectrum for one instance of disorder appears very random and its average is featureless. Notably it was checked numerically that for the same type of disorder used for the Weyl semimetal on a 3D strong topological insulator indeed removed all delta-function sharp signatures from $C(E,E';\vec{k},\vec{k}')$. Thus the sharp picks discussed here and there are of a different origin and character. Second, here the interest is mainly in correlations between different energies ($P(\omega \neq 0,q)$). It would be interesting to check whether time dependent STM measurements may give a proxy to $P(\omega,q)$ with non-zero $\omega$ as this may open the possibility of measuring such correlations using STM as well as ARPES.

\end{document}